\documentclass[aps,prd,preprint,showpacs]{revtex4}
\usepackage{graphicx}
\usepackage{subfigure}
\usepackage{hyperref}
\usepackage{amsmath}
\begin{document}
\draft
\title{Anomalous quartic $WW\gamma\gamma$ and
$ZZ\gamma\gamma$ couplings in $\gamma p$ collision at the LHC}

\author{A. Senol}\email{asenol@kastamonu.edu.tr}
\affiliation{Kastamonu University, Department of Physics, 37100,
Kastamonu, Turkey}
\begin{abstract}
We analyze the anomalous quartic gauge boson couplings
$WW\gamma\gamma$ and $ZZ\gamma\gamma$, described by dimension-6
effective quartic Lagrangian at the LHC. The sensitivities to
anomalous quartic gauge couplings $a_{0,c}^{W,Z}/\Lambda^2$ by
examining the two different photon-induced processes $pp\to p\gamma
p\to p W\gamma qX$ and $pp\to p\gamma p\to p ZZ qX$  with $W$ and
$Z$s decaying leptonically are investigated. We show that $\gamma p$
mode of photon-induced reactions at the LHC are able to probe these
couplings to the order of $10^{-6}$-$10^{-7}$ GeV$^{-2}$ at 95\%
confidence level with $\sqrt s$ = 14 TeV and for proton-proton
luminosities in the range of 30-200 fb$^{-1}$.
\end{abstract}
 \maketitle
\section{introduction}
The structure of triple and quartic interactions of the gauge bosons
in the electroweak sector of the Standard Model (SM) are represented
by the non-Abelian $SU(2)_L\times U(1)_Y$ local gauge symmetry.
Possible deviations of the triple and quartic gauge boson couplings
from SM predictions within the experimental precision can give
valuable information about new physics beyond the SM. A simple way
to parameterize these new physics effects at higher energies is to
assume that the SM is an effective theory at low energies. Genuine
quartic gauge couplings arise from effective operators which do not
lead to any trilinear gauge boson couplings. Therefore, the SM can
be extended via the trilinear gauge couplings that are equal to
their SM values while quartic gauge couplings are modified by
genuine anomalous interactions. In this way, quartic gauge boson
couplings can be constrained independently of the bounds on the
anomalous trilinear vertices. The two independent C and P conserving
dimension-6 effective quartic Lagrangian operators involving at
least two photons that give rise to genuine anomalous quartic
couplings imposing local $U(1)_{EM}$ and custodial $SU(2)_{Weak}$
symmetry are \cite{Belanger:1992qh,Belanger:1992qi}
\begin{eqnarray}\label{eq1}
{\cal
L}_{0}&&=\frac{-e^2}{8}\frac{a_{0}^W}{\Lambda^{2}}F_{\mu\nu}F^{\mu\nu}
W^{+\alpha}W^-_{\alpha}-\frac{e^2}{16\cos^2\theta_W}\frac{a_{0}^Z}{\Lambda^{2}}F_{\mu\nu}F^{\mu\nu}
Z^{\alpha}Z_{\alpha}
\end{eqnarray}
and
\begin{eqnarray}\label{eq2}
{\cal
L}_{c}&&=\frac{-e^2}{16}\frac{a_{c}^W}{\Lambda^{2}}F_{\mu\alpha}F^{\mu\beta}
(W^{+\alpha}W_{\beta}^{-} +
W^{-\alpha}W_{\beta}^{+})-\frac{e^2}{16\cos^2\theta_W}\frac{a_{c}^Z}{\Lambda^{2}}F_{\mu\alpha}F^{\mu\beta}
Z^{\alpha}Z_{\beta},
\end{eqnarray}
where $W^{\pm}_\alpha$ is the $W^{\pm}$ boson field, $F_{\mu\nu}$ is
the tensor for electromagnetic field strength, $a_0^{W(Z)}$ and
$a_c^{W(Z)}$ are the dimensionless anomalous coupling constants of W
(Z) parts of the Lagrangian, and $\Lambda$ is interpreted as the
energy scale of the new physics. The anomalous couplings are zero in
the SM.

The interaction Lagrangians ${\cal L}_{0}$ and ${\cal L}_{c}$
generate anomalous contributions to two $WW\gamma\gamma$ vertices as
given by \cite{Eboli:1993wg}
\begin{eqnarray}\label{vertex1}
i\frac{2\pi\alpha}{\Lambda^{2}}a_{0}^Wg_{\mu\nu}\left[g_{\alpha\beta}
(p_{1}.p_{2})-p_{2 \alpha}p_{1 \beta}\right]
\end{eqnarray}
and
\begin{eqnarray}\label{vertex2}
i\frac{\pi\alpha}{2\Lambda^{2}}a_{c}^W\left[(p_{1}.p_{2})(g_{\mu\alpha}
g_{\nu\beta}+g_{\mu\beta}g_{\alpha\nu})+g_{\alpha\beta}(p_{1\mu}p_{2\nu}
+p_{2\mu}p_{1\nu})\right. \nonumber \\
\left. -p_{1\beta}(g_{\alpha\mu}p_{2\nu}+g_{\alpha\nu}p_{2\mu})
-p_{2\alpha}(g_{\beta\mu}p_{1\nu}+g_{\beta\nu}p_{1\mu})\right],
\end{eqnarray}
where the fine structure constant is $\alpha=e^2/(4\pi)$, $p_{1}$
and $p_{2}$ are the four-momenta of photons. The anomalous
$ZZ\gamma\gamma$ vertex is derived by multiplying above vertex
functions Eq. (\ref{vertex1}) and Eq. (\ref{vertex2}) by
$1/\cos^2\theta_W$ and with the replacement $W \to Z$. The
$ZZ\gamma\gamma$ vertex does not occur in SM at tree level.

All anomalous couplings in the effective Lagrangian Eqs. (\ref{eq1})
and (\ref{eq2}) cause tree-level unitarity violation at high
energies. The standard procedure to regularise the cross section is
to employ a dipole form factor:
\begin{eqnarray}
a_{0,c}^{W,Z}(\hat{s})=\frac{a_{0,c}^{W,Z}}{(1+\hat{s}/\Lambda_{cutoff}^2)^2}
\end{eqnarray}
where $\hat{s}$ is the partonic center of mass energy and
$\Lambda_{cutoff}$ is the scale of new physics. We obtained the
limits on anomalous couplings to compare our results with the
scenario $\Lambda_{cutoff}\to \infty$. The maximal
$\Lambda_{cutoff}$ is calculated from the given value of anomalous
couplings which can be in form factors. To protect the unitarity, in
this study we calculated the maximal $\Lambda_{cutoff}$ to be about
3 TeV when order of $10^{-7}$ is taken for
$a_{0,c}^{W,Z}/\Lambda^2$.

The anomalous $a_{0}^{W,Z}/\Lambda^{2}$ and
$a_{c}^{W,Z}/\Lambda^{2}$ couplings were experimentally limited at
the 95\% C.L. by the OPAL collaboration from measurements of
$WW\gamma$, $q\bar q\gamma\gamma$, and $\nu\bar \nu\gamma\gamma$
production at CERN LEP collider \cite{Abbiendi:2004bf}. Recently,
the experimental 95\% C.L. limits on anomalous couplings
$a_{0,c}^{W}$ have been provided by the D0 \cite{Abazov:2013opa}
collaboration at the Fermilab Tevatron from events with dielectron
and missing energy, and by the CMS \cite{Chatrchyan:2013foa}
collaboration at CERN LHC from exclusive two-photon production of
$W^+W^-$. All limits are given in Table \ref{limits}.
\begin{table}
\caption{The 95\% C.L. upper limits on anomalous quartic
$WW\gamma\gamma$ and $ZZ\gamma\gamma$ couplings without form
factors.\label{limits}}
 \small{\begin{tabular}{lccccc}
  \hline
  Parameters  [$\textmd{GeV}^{-2}$] & CMS                   && D0                     && OPAL
  \\\hline

  $a_0^{W}/\Lambda^{2}$ & [-4.0$\times10^{-6}$; 4.0$\times10^{-6}$]&& [-0.00043, 0.00043] && [-0.020, -0.020]  \\
  $a_c^W/\Lambda^{2} $   & [-1.5$\times10^{-5}$,1.5$\times10^{-5}$] && [-0.0015, 0.0015]   && [-0.052, 0.037] \\
  $a_0^{Z}/\Lambda^{2}$ & - && - && [-0.007, 0.023] \\
  $a_c^Z\Lambda^{2}$   & - && - && [-0.029, 0.029]\\
  \hline
\end{tabular}}
\end{table}

LHC will allow probing of new physics via photon-induced
interactions at energies beyond the electroweak energy scale by
allowing the use of complementary information to the parton-parton
collisions at the LHC by adding forward proton detectors
\cite{deFavereaudeJeneret:2009db}. For instance, the use of forward
proton tagging for the measurements of outgoing scattered proton
momenta, would provide spin-parity information about exclusively
produced particles in the photon-induced processes
\cite{Heinemeyer:2007tu}. Photon-induced processes include a low
virtuality quasi-real photon which is scattered with small angle
from the beam pipe. Therefore, the photon emitting intact proton is
scattered with small angle and thus escapes from the central
detectors of CMS and ATLAS without being detected. This intact
scattered protons in the final state leave a characteristic sign in
the forward detectors which are suggested to be located at distances
of 220 m and 420 m from the interaction point according to the
forward physics program of CMS and ATLAS collaborations
\cite{Albrow:2010yb, Tasevsky:2009zza, Tasevsky:2014cpa}. The
photon-induced reactions provide a suitable platform of searching
for photonic-quartic anomalous gauge couplings thanks to these
distinctive experimental features. At the LHC, photonic quartic
$WW\gamma\gamma$ and $ZZ\gamma\gamma$ vertices are probed in
photon-induced reactions, i.e. $pp\to p\gamma\gamma p\to p W^+W^- p$
\cite{Pierzchala:2008xc,Chapon:2009hh} for $WW\gamma\gamma$
couplings and $pp\to p\gamma\gamma p\to p Z Z p$
\cite{Chapon:2009hh,Gupta:2011be}, $pp\to p\gamma p\to p\gamma q Z
X$ \cite{Sahin:2012mz} for $ZZ\gamma\gamma$ couplings which were
elaborately studied in the literature. Furthermore, particularly
well suited phenomenological studies of anomalous vertices
$WW\gamma\gamma$ and $ZZ\gamma\gamma$ have already been performed at
the LHC via traditional pp reactions \cite{Belyaev:1998ih,
Dervan:1999as,Eboli:2000ad,Eboli:2003nq,Eboli:2006wa,Yang:2012vv},
$e^+e^-$ colliders
\cite{Belanger:1992qh,AbuLeil:1994jg,Han:1997ht,Boos:1997gw,Boos:1999kj,Stirling:1999ek,Belanger:1999aw,Denner:2001vr,Montagna:2001ej,Gutierrez-Rodriguez:2013eya}
 and its $\gamma\gamma$
\cite{Belanger:1992qi,Eboli:1995gv,Eboli:2001nb,Sahin:2008ej},
$e\gamma$ \cite{Eboli:1993wg,Atag:2007ct} modes. In this work, we
study the anomalous quartic gauge boson couplings $WW\gamma\gamma$
and $ZZ\gamma\gamma$ by examining the two different photon-induced
processes which are $pp\to p\gamma p\to p W\gamma qX$ and $pp\to
p\gamma p\to pZZ qX$ at the LHC.
\begin{figure*}[htbp!]
  \includegraphics[width=16cm]{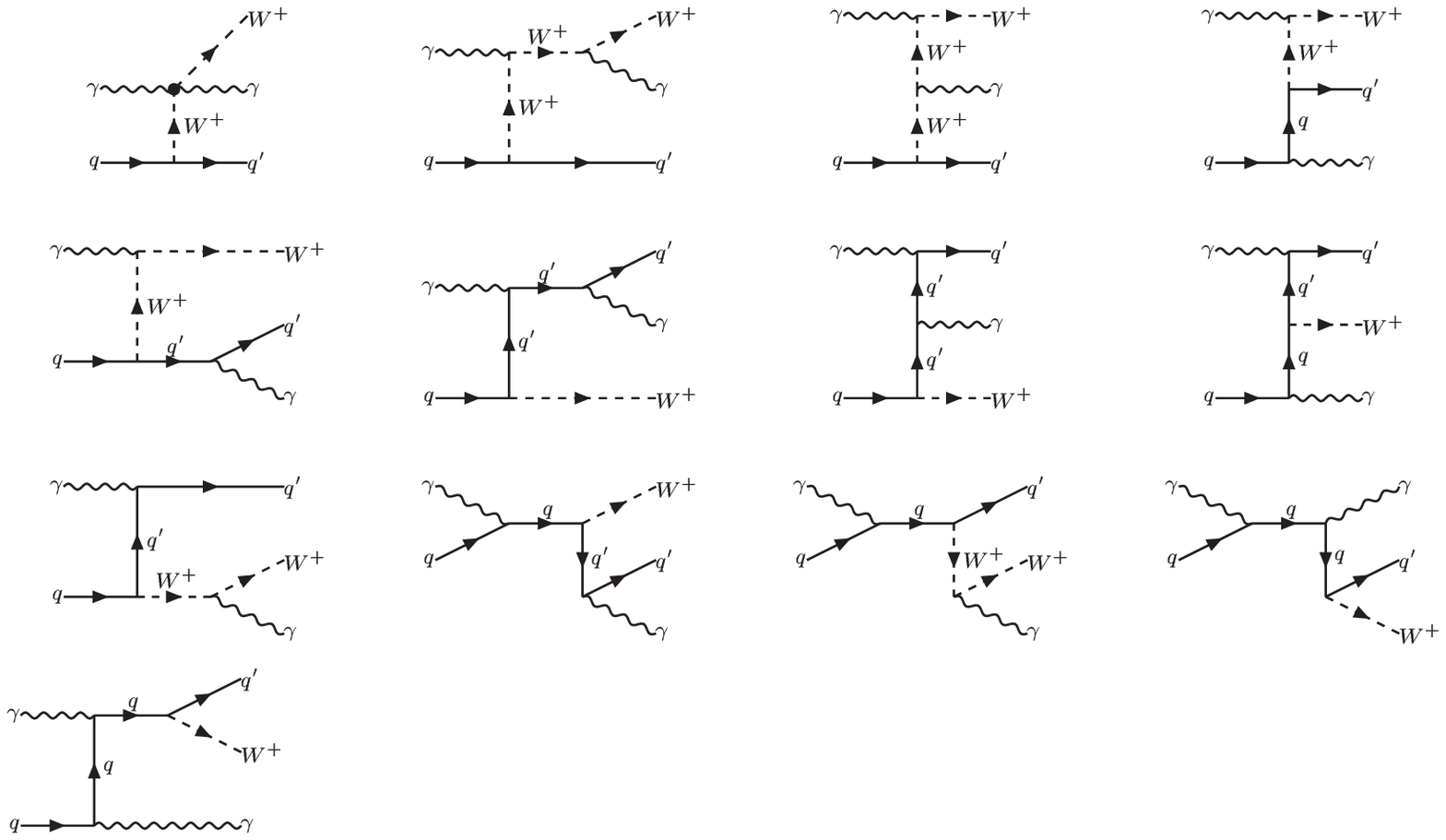}\\
 \caption{Feynman graphs for the tree-level subprocess $\gamma q\to W \gamma q'$ (where $q=u,c,\bar d,\bar s$ and
$q'=d,s,\bar u,\bar c$).}\label{fd1}
\end{figure*}
\begin{figure*}[htbp!]
  \includegraphics[width=14cm]{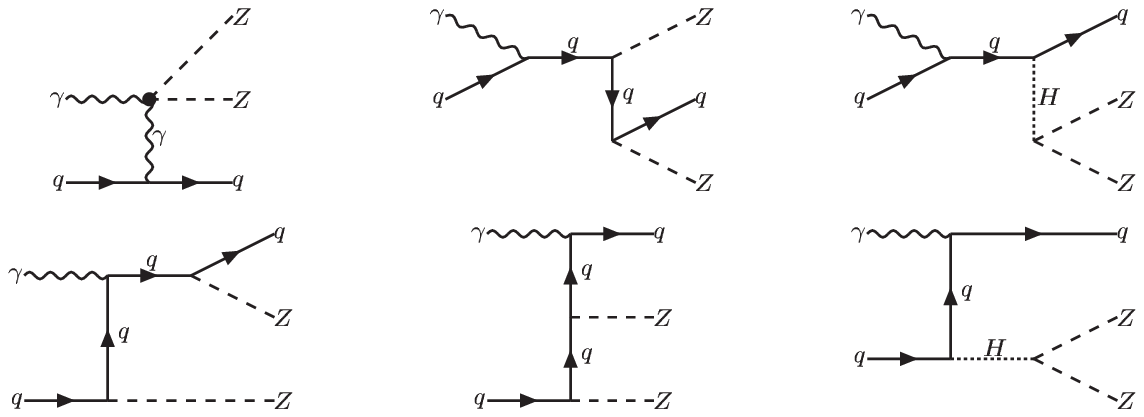}\\
 \caption{Feynman graphs for the tree-level subprocess $\gamma q\to ZZ q$ (where $q=u,\bar u,d,
\bar d, c,\bar c,s,\bar s, b,\bar b$.}\label{fd2}
\end{figure*}

\section{The cross sections for the production of $W\gamma$ and $ZZ$ in $\gamma p$ collision }
The tree-level SM Feynman diagrams of the subprocess $\gamma q\to W
\gamma q'$ in the main reaction $pp\to p\gamma p\to p W\gamma qX$
are shown in Fig. \ref{fd1}. The first of these diagrams receive
contributions from the anomalous $WW\gamma\gamma$ couplings. In the
case of examining anomalous $ZZ\gamma\gamma$ couplings, we consider
the subprocess $\gamma q\to ZZ q$ of the main reaction $pp\to
p\gamma p\to p ZZ qX$. The anomalous $ZZ\gamma\gamma$ vertex
contributions are shown in the first diagrams of Fig. \ref{fd2},
whereas the others depict the tree-level SM Feynman diagrams. All
calculations were evaluated using the tree-level event generator
CalcHEP \cite{Belyaev:2012qa}, by adding the vertex functions Eqs.
(\ref{vertex1}) and (\ref{vertex2}). The total cross sections for
$pp\to p\gamma p\to p W\gamma qX$ and $pp\to p\gamma p\to p ZZ qX$
processes can be obtained by integrating the cross sections for the
subprocess $\gamma q\to W \gamma q'$ (where $q=u,c,\bar d,\bar s$
and $q'=d,s,\bar u,\bar c$) and $\gamma q\to ZZ q$ (where $q=u,\bar
u,d, \bar d, c,\bar c,s,\bar s, b,\bar b$) with the photon and quark
distributions:
\begin{eqnarray}
\sigma\left(
\begin{array}{c}
  pp\to p\gamma p\to p W \gamma q X \\
   pp\to p\gamma p\to p Z Z q X
\end{array}\right)=\int_{Q^{2}_{min}}^{Q^{2}_{max}} {dQ^{2}}\int_{x_{1\;
min}}^{x_{1\;max}} {dx_1 }\int_{x_{2\; min}}^{x_{2\;max}} {dx_2}
\left(\frac{dN_\gamma}{dx_1dQ^{2}}\right)\nonumber\\
\times\left(\frac{dN_q}{dx_2}\right)\hat{\sigma}\left(\begin{array}{c}
\gamma q\to W\gamma q \\
\gamma q \to Z Z q\end{array}\right)(\hat s)
\end{eqnarray}
where $x_1=\frac{E_\gamma}{E}$ (here $E$ denotes the energy of the
incoming proton beam and  $E_{\gamma}$ is the photon energy), $x_2$
is the momentum fraction of the proton's momentum carried by the
quark, $\frac{dN_q}{dx_2}$ is the quark distribution function of the
proton and $\frac{dN_\gamma}{dx_1dQ^{2}}$ is the photon spectrum in
equivalent photon approximation (EPA). In numerical calculations, we
use CTEQ6L \cite{Pumplin:2002vw} for parton distribution functions
and the EPA \cite{Budnev:1974de,Ginzburg:1981vm,Piotrzkowski:2000rx}
embedded in CalcHEP for the photon spectra. The photon spectrum of
virtuality $Q^2$ and energy $E_{\gamma}$ in EPA is defined by the
following formula \cite{Budnev:1974de,Piotrzkowski:2000rx}:
\begin{eqnarray}
\label{spectrum1}
\frac{dN_\gamma}{dE_{\gamma}dQ^{2}}=\frac{\alpha}{\pi}\frac{1}{E_{\gamma}Q^{2}}
[(1-\frac{E_{\gamma}}{E})
(1-\frac{Q^{2}_{min}}{Q^{2}})F_{E}+\frac{E^{2}_{\gamma}}{2E^{2}}F_{M}]
\end{eqnarray}
where $Q^{2}_{min}$ denotes the photon minimum virtuality is given
by
\begin{eqnarray*}
&&Q^{2}_{min}=\frac{m^{2}_{p}E^{2}_{\gamma}}{E(E-E_{\gamma})}
\end{eqnarray*}
here, $m_p$ is the mass of the incoming proton. The magnetic and
electric form factors $F_{M}$ and $F_{E}$ are defined by
\begin{eqnarray*}
&&F_{E}=\frac{4m^{2}_{p}G^{2}_{E}+Q^{2}G^{2}_{M}}
{4m^{2}_{p}+Q^{2}},\;\; \;\;\; F_{M}=G^{2}_{M}\\
&&G^{2}_{E}=\frac{G^{2}_{M}}{7.78}=(1+\frac{Q^{2}}{0.71\mbox{GeV}^{2}})^{-4}
\end{eqnarray*}
In our calculations, we have taken $Q^2_{max}$=2 GeV$^2$ for which
the contribution to the integral above this value is very small.

\begin{figure*}[htbp!]
  \includegraphics[width=12cm]{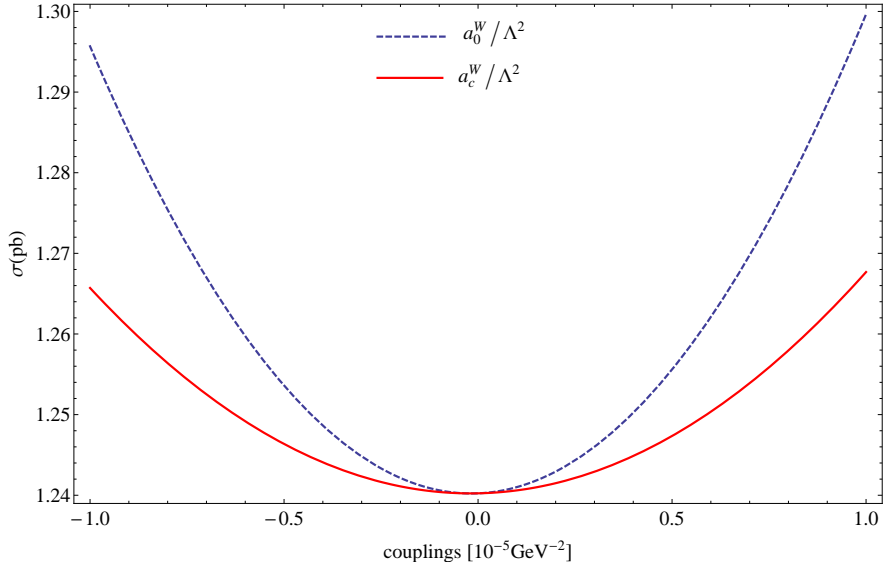}\\
 \caption{The total cross sections depending on anomalous $a_{0}^{W}/\Lambda^2$ and $a_c^W/\Lambda^2$ couplings for the process $pp\to p\gamma p\to p W\gamma
qX$ at the LHC with $\sqrt s$= 14 TeV.}\label{fig1}
\end{figure*}

\begin{figure*}[htbp!]
  \includegraphics[width=12cm]{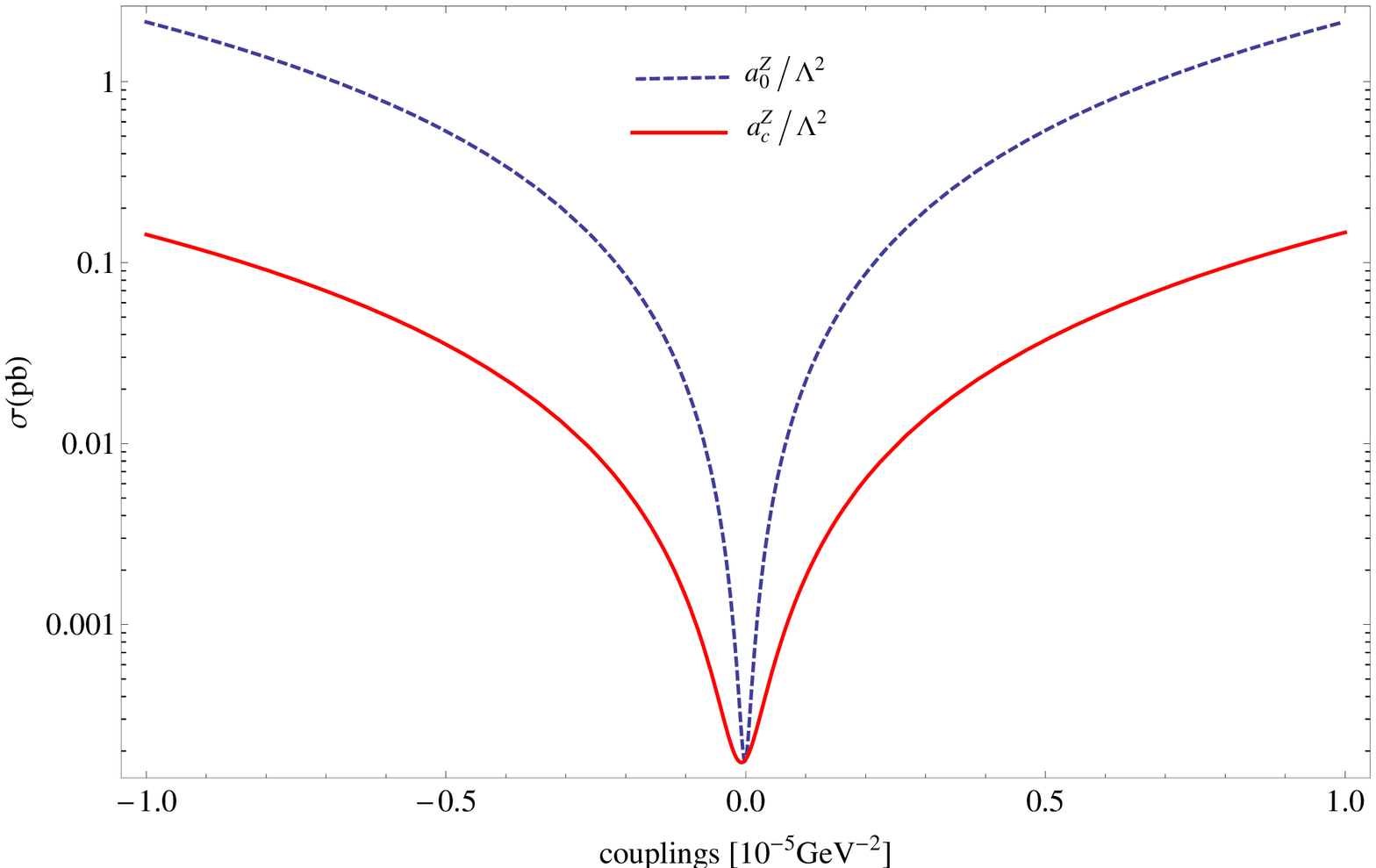}\\
 \caption{The total cross sections as function of anomalous $a_{0}^{Z}/\Lambda^2$ and $a_c^Z/\Lambda^2$ couplings for the process $pp\to p\gamma p\to p ZZ qX$ at the LHC with $\sqrt s$= 14 TeV.}\label{fig2}
\end{figure*}

The total cross sections of the processes $pp\to p\gamma p\to p
W\gamma qX$ and $pp\to p\gamma p\to p ZZ qX$ are given in Fig.
\ref{fig1} and Fig. \ref{fig2} as functions of anomalous
$a_{0,c}^{W}/\Lambda^2$ and $a_{0,c}^{Z}/\Lambda^2$ couplings at the
LHC with $\sqrt s=$ 14 TeV. In these figures, the cross sections
depending on the anomalous quartic gauge coupling parameter were
obtained by varying only one of the anomalous couplings at a time
while the other was fixed to zero.

\section{Sensitivity to the anomalous $WW\gamma\gamma$ and $ZZ\gamma\gamma$ couplings}
The bounds of anomalous $a_0^{W,Z}/\Lambda^2$ and
$a_c^{W,Z}/\Lambda^2$ couplings at 95\% C.L. were obtained by
applying one and two-dimensional $\chi^2$ tests without considering
systematic errors. $\chi^2$ is defined as:
\begin{eqnarray}
\chi^{2}=\left(\frac{\sigma_{SM}-\sigma_{AN}}{\sigma_{SM} \,\,
\delta}\right)^{2}
\end{eqnarray}
where $\sigma_{AN}$ is the cross section in the presence of
anomalous couplings, $\delta=\frac{1}{\sqrt{N}}$ is the statistical
error and here $N$ is the number of events. The number of events for
$pp\to p\gamma p\to p W\gamma qX$ is given by $N=E\times
S\times\sigma_{SM}\times L_{int}\times BR(W\to l \nu)$ where $E$ is
the jet reconstruction efficiency, $S$ denotes the survival
probability factor, $\sigma_{SM}$ is the corresponding SM background
cross section, $L_{int}$ is the integrated luminosity and $l=e^-$ or
$\mu^-$. Similarly, for $pp\to p\gamma p\to p ZZ qX$ process
$N=S\times E\times\sigma_{SM}\times L_{int}\times BR(Z\to l \bar
l)^2$. We also assume $S=0.7$ and $E=0.6$ for both processes, as in
Ref. \cite{Khoze:2002dc,deFavereaudeJeneret:2009db,Sahin:2012mz}. A
$p_T^{j,\gamma}> 15$ GeV cut was applied on the transverse momenta
of final state photons and jets. We also imposed the pseudorapidity
cuts $|\eta^{j,\gamma}|< 2.5$ on final state photons and jets
because pseudorapidity coverage of central detectors of CMS and
ATLAS is $|\eta|< 2.5$. We do not consider any acceptance for the
final state leptons because our calculations do not provide the
lepton momenta.

To discern the photoproduction process from the usual proton-proton
backgrounds and close the intrinsic $p_T$ spread of the LHC beams,
we apply a $p_T>100$ MeV cut on the transverse momentum of outgoing
protons that emit photons
\cite{Albrow:2010yb,Piotrzkowski:2000rx,Cox:2005if}.

The calculated one-dimensional limits (with the other anomalous
coupling fixed to zero) on anomalous quartic gauge couplings
$a_{c}^{W,Z}/\Lambda^{2}$ and $a_{0}^{W,Z}/\Lambda^{2}$ at 95\% C.L.
sensitivity for some integrated luminosities are given in Table
\ref{1D}.

\begin{table}
\caption{95\% C.L. constrains on anomalous quartic gauge couplings
$WW\gamma\gamma$ and $ZZ\gamma\gamma$ parameters
$a_{c}^{W,Z}/\Lambda^{2}$ and $a_{0}^{W,Z}/\Lambda^{2}$ at LHC with
$\sqrt s$= 14 TeV.\label{1D}  }
\begin{tabular}{ccccc}
  \hline
  L(fb$^{-1}$) & $a_0^W/\Lambda^2(\times 10^{-6}$ GeV$^{-2}$) &  $a_c^W/\Lambda^2(\times
  10^{-6}$ GeV$^{-2}$)&$a_0^Z/\Lambda^2(\times 10^{-7}$ GeV$^{-2}$)&$a_c^Z/\Lambda^2(\times 10^{-7}$ GeV$^{-2}$)
  \\\hline
  30 & [-8.67; 8.32] & [-12.71; 12.33]& [-5.82; 5.58] &[-22.62; 21.24] \\
  50 & [-7.65; 7.31] & [-11.21; 10.83] & [-5.13; 4.90]& [-19.99; 18.61]\\
  100 & [-6.46; 6.12] & [-9.45; 9.08] & [-4.33; 4.10]&[-16.93; 15.54] \\
  200 & [-5.46; 5.12] & [-7.98; 7.61] & [-3.66; 3.43]&[-14.35; 12.97] \\
  \hline
\end{tabular}
\end{table}

Our obtained limits on $a_0^{W,Z}/\Lambda^2$ and
$a_c^{W,Z}/\Lambda^2$ are approximately four orders of magnitude
more restrictive than the best limits obtained from OPAL
\cite{Abbiendi:2004bf} as can be seen from the comparison of Table
\ref{limits} and Table \ref{1D} . On the other hand, the bounds for
$a_0^{W}/\Lambda^2$ and $a_c^{W}/\Lambda^2$ from D0 collaboration at
Tevatron \cite{Abazov:2013opa} are worse than our values by a factor
of order two while the CMS limits at $\sqrt s$=7 TeV with
$L_{int}$=5 fb$^{-1}$ \cite{Chatrchyan:2013foa} have similar
sensitivity as our limits.

\begin{figure*}[htbp!]
  \includegraphics[width=12cm]{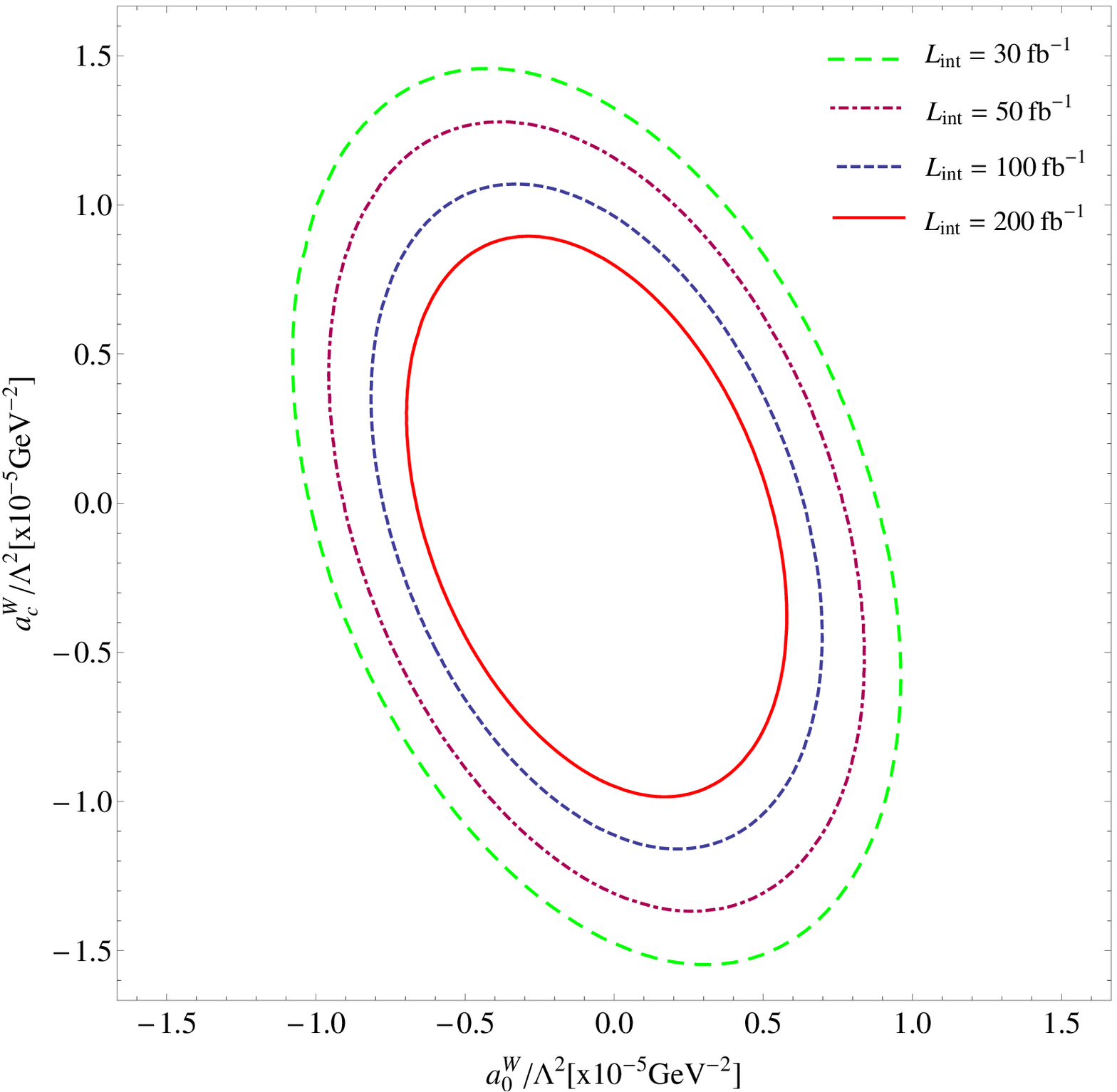}\\
 \caption{95\% C.L. contours for anomalous $a_0^{W}/\Lambda^2$ and
$a_c^{W}/\Lambda^2$ couplings for the process $pp\to p\gamma p\to p
W\gamma qX$ at the LHC with $\sqrt s$= 14 TeV.}\label{cont1}
\end{figure*}

\begin{figure*}[htbp!]
  \includegraphics[width=12cm]{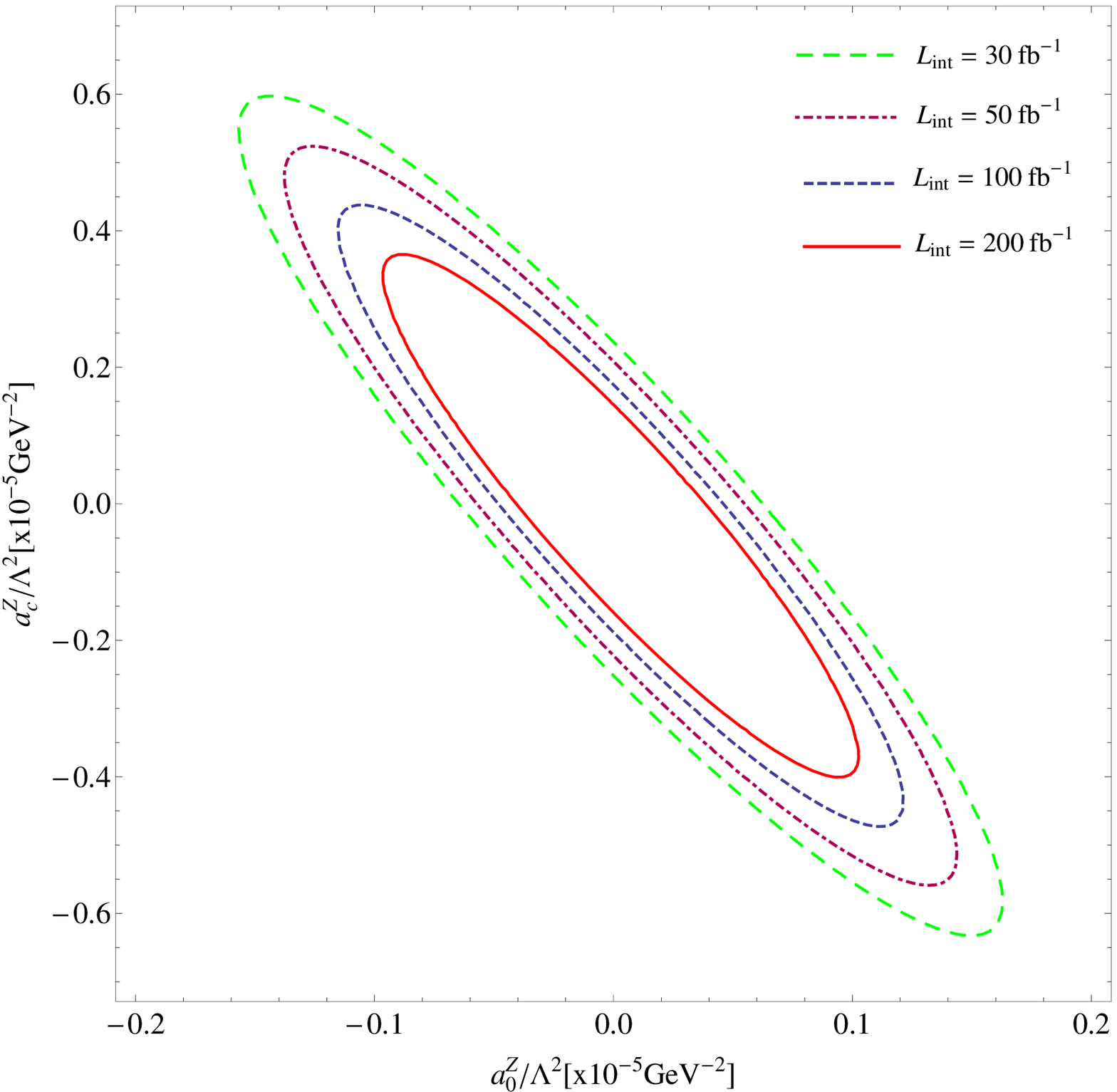}\\
 \caption{95\% C.L. contours for anomalous $a_0^{Z}/\Lambda^2$ and
$a_c^{Z}/\Lambda^2$ couplings for the process  $pp\to p\gamma p\to p
ZZ qX$ at the LHC with $\sqrt s$= 14 TeV.}\label{cont2}
\end{figure*}

In addition, we present 95\% C.L. contours in the
$a_0^{W}/\Lambda^2$-$a_c^{W}/\Lambda^2$ plane in Fig.\ref{cont1} and
the $a_0^{Z}/\Lambda^2$-$a_c^{Z}/\Lambda^2$ plane in Fig.\ref{cont2}
at $\sqrt s$=14 TeV for various integrated luminosities. As we can
see from Fig.\ref{cont1}, the best limits on $a_0^{W}/\Lambda^2$ and
$a_c^{W}/\Lambda^2$ through the reactions $pp\to p\gamma p\to p
W\gamma qX$ are [$-6.5\times 10^{-6}; 6.0\times 10^{-6}$] GeV$^{-2}$
and [$-9.5\times 10^{-6}; 8.5\times 10^{-6}$] GeV$^{-2}$,
respectively for L$_{int}$=200 fb$^{-1}$ at the LHC. According to
Fig.\ref{cont2}, the attainable bounds on $a_0^{Z}/\Lambda^2$ and
$a_c^{Z}/\Lambda^2$ via reactions $pp\to p\gamma p\to p ZZ qX$ are
[$-1.0\times 10^{-6}; 1.0\times 10^{-6}$] GeV$^{-2}$ and
[$-4.0\times 10^{-6}; 3.5\times 10^{-6}$] GeV$^{-2}$, respectively.
\section{conclusions}
The high energy photon-photon or photon-proton interactions at the
LHC exhibit a suitable platform to probe genuine anomalous quartic
gauge couplings. Especially, the photon-photon reactions can provide
much higher sensitivity than partonic reactions due to clean
experimental conditions and mostly free from QCD backgrounds for
anomalous quartic gauge couplings. On the other hand, photon-proton
reactions have higher luminosities and higher center of mass
energies compared to photon-photon reactions. Since the anomalous
quartic gauge boson couplings involve higher luminosity and higher
center of mass energies, it is more proper to study them in
photon-proton reactions. In this work, we have performed an analysis
of the $pp\to p\gamma p\to p W\gamma qX$ and $pp\to p\gamma p\to p
ZZ qX$ processes with $W$ and $Z$s decaying leptonically in order to
assess the sensitivities to anomalous quartic gauge couplings
$a_{0,c}^{W,Z}/\Lambda^2$ by using dimension-6 effective quartic
Lagrangian at LHC assuming triple gauge boson couplings $WW\gamma$
to be at their SM values. We showed that our limits are several
orders of magnitude beyond the best limits obtained from LEP
\cite{Abbiendi:2004bf} and Tevatron \cite{Abazov:2013opa}. Our
limits have similar sensitivity with those obtained from CMS
\cite{Chatrchyan:2013foa} at $\sqrt s$=7 TeV with $L_{int}$=5
fb$^{-1}$. The results of $pp\to p\gamma p\to p W\gamma qX$ and
$pp\to p\gamma p\to p ZZ qX$ processes in our study  are less
sensitive than the results of ref. \cite{Pierzchala:2008xc} which
are obtained by the fully exclusive production.
\begin{acknowledgements}
I would like to thank Abant Izzet Baysal University Department of
Physics where this study was carried out, for their hospitality and
Orhan Cakir for useful comments and discussions.
\end{acknowledgements}
  
\end{document}